\documentstyle[12pt,aas2pp4]{article}

\begin{document}

\title{A New Model for the Spiral Structure of the Galaxy. \\ 
Superposition of 2+4-armed patterns}
\author{J.R.D.L\'epine}
\affil{ Instituto Astron\^omico e Geofisico da USP, Av.Miguel
Stefano 4200,\\
04301-904 S\~ao Paulo, SP, Brazil; E-mail: jacques@radio.iagusp.usp.br}
\and
\author{Yu.N.Mishurov, S.Yu.Dedikov}
\affil{ Space Research Department, Rostov State    
University, 5 Zorge, Rostov-on-Don,\\
344090, Russia; E-mail: mishurov@phys.rnd.runnet.ru}

\begin {abstract} 

We investigate the possibility of describing the spiral pattern of the
Milky Way  in terms of a  model of superposition 2- and 4-armed wave
harmonics (the simplest description, besides pure modes). Two
complementary methods are used: a study of stellar kinematics,
and direct tracing of positions of spiral arms. In the first method,
the parameters of the galactic rotation curve and the free parameters of the
spiral density waves were obtained from Cepheid kinematics, under different
assumptions. To turn visible the structure corresponding to these models, 
we computed  the evolution of an  ensemble of $N$-particles, simulating the
ISM clouds, in the perturbed galactic gravitational field.
In the second method, we present a new analysis of the longitude-velocity
($l-v$) diagram of the sample of galactic HII regions, converting positions of
spiral arms in the galactic plane into locii of these arms in the $l-v$
diagram. Both methods indicate that the ``self-sustained'' model, in which the
2-armed and 4-armed mode have different pitch angles (6$\arcdeg$ and 
12$\arcdeg$, respectively) is a good description of the disk structure.
An important conclusion is that the Sun happens to be practically at the
corotation circle. As an additional result of our study, we  propose an
independent test for localization of the corotation circle in a spiral
galaxy: a gap in the radial distribution of interstellar gas has to be
observed in the  corotation region.

\end{abstract}


\section{Introduction} 

A good understanding of  the large-scale spiral structure of
the  Galaxy  has not been reached up to the present. Georgelin \& 
Georgelin\markcite{GG} (1976), herafter GG,  derived a  4-armed pattern,
 based on an analysis of the distribution of giant H II regions.
 According to Vall\'ee \markcite{Va}
(1995) most researches support the 4-armed pattern, although there 
are discordant opinions; for instance Bash \markcite{Ba}(1981) finds
the pattern to be 2-armed, similar to the first spiral structure model
proposed by Lin and Shu \markcite{LS} (1964). But even if we accept that
there are strong observational evidences in favor of a 4-arms structure,
theoretical and observational difficulties remain, and the observation
of external galaxies strongly recommends that we look for more complex
solutions, which should include  arms with different pitch angles.

It is now accepted that the distance to the center is about 7.5 kpc
(eg. Olling \& Merrefield,\markcite{OM} 1998), which implies a smaller
 rotation velocity of the Local Standard of Rest than previously
 estimated, and as a consequence, the rotation curve  decreases
 beyond the solar radius (Amaral et al.,\markcite{AOL} 1996, Honma \& Kan-Ya,
 \markcite{HK} 1998). From the rotation curve, the epicyclic frequency
 and the radius of the inner and outer Lindblad resonances can be derived.
 Amaral and L\'epine\markcite{AL} (1997, hereafter AL)
concluded that the 2-armed pattern could exist between 2.5 and 12 kpc, 
which is about the range of the observed spiral pattern, but a
4-armed pattern could only exist between about 6 to 11 kpc. AL proposed
a representation of the galactic spiral structure as a superposition of
2- and 4-armed wave patterns, to account for the number of arms and for the
existence of arms over a wider range of radius. In AL's model, two arms
of the 4-armed component are coincident with the 2-armed component, so
that the Galaxy looks 4-armed. However, there are theoretical arguments 
discussed in the present paper showing that the pitch angle for the 4-armed 
pattern should be different from that of the 2-armed component. A Fourier
analysis of external galaxies in terms of spiral modes, performed by 
Puerari \& Dottori \markcite{PD} (1992) indicates as well that in the
cases in which the 2-armed and 4-armed patterns are prominent, they indeed
 have different pitch angles. These arguments suggest that at least
in a range of radius, the Galaxy might present a structure like 2-arms
plus 4-arms with  different pitch angles, which could look like 6 arms.
An attractive aspect of such a model with arms of different pitch
angles is that it naturally accounts for bifurcations of arms. In
external galaxies branching of arms is a widespread phenomenon, e.g.
the galaxy M 101, where there are both branching arms and bridges
between them. The structure of the Milky Way is perhaps  similarly
complicated.  Another fact that stimulates us to look for more refined
 solutions to the spiral structure is that there are observed arms in
 the Galaxy that do not fit in a simple 4-arms structure. For instance,
 outside the solar circle, around longitudes 210$\arcdeg$ - 260$\arcdeg$,
 three arms are clearly visible, in
HI and in IRAS sources (Kerr, 1969, Wouterloot et al., 1990), but they are not
expected from a 4-arms model that fits the tangential directions of the
inner parts of the Galaxy (e.g. Ortiz \& L\'epine, 1993).

The main goal of this paper is to investigate if a superposition of 
2+4-armed wave harmonics is a good representation of  the spiral
structure of the Galaxy. Two different approaches are used. In one 
approach, we analyse  the Cepheid kinematics, using a model that takes
into consideration the  perturbation of the stellar velocity field by the
spiral arms, to derive the structural parameters for the gravitational
field of the galactic disk. And, since the structure of the perturbation
potential does not necessarily correspond to the visible structure,
we performed a many-particle simulation of gas dynamics in this potential,
to predict the visible structure.
In the second approach, we directly analyse the visible structure
by means of the  sample of HII regions, since these objects are
recognized as the best large-scale tracers of the galactic structure. We
 present a new analysis of the observed longitude-velocity ($l-v$) diagram
 of the best up-to-date sample of  H II regions,
deriving from it the position of the spiral arms, and we compare the
observed $l-v$ diagram with the theoretical ones, computed from our
gas dynamics simulations.
  
The Cepheid kinematics analysis is performed for two different
models of 2-arms + 4-arms structures, and the results are compared with
previous calculations of pure 2-arms and pure 4-arms. An important result
is that for the more reallistic model, the Sun lies very
close to the corotation region. Furthermore, our simulation  revealed 
the interesting effect of gas pumping out from the corotation. This suggests
that in  spiral galaxies, there must be a gas deficiency  in a region near the
corotation circle. This phenomenon explains the lack of atomic hydrogen in
a ring-like region near the solar circle derived by Kerr (1969) and by
Burton (1976), which remained ununderstood up to now. We propose this effect
as an independent test for localization of a corotation circle in 
external spiral galaxies.


\section{Method of estimation of structural parameters}
 
The  derivation of the spiral wave parameters is based on the
statistical analysis of stellar motion in the Galaxy (see Cr\'ez\'e \&
Mennessier, 1973,  Mishurov \& Zenina, 1999, hereafter MZ, Mishurov et
al., 1997, hereafter MZDMR and papers cited therein). Note that we look for
the parameters of the structure of the galactic gravitational field, which
 is not directly visible. However, the gravitational field
determines the stellar motion, and in particular, the spiral perturbations of
the field deviate the stellar motion from rotation symmetry. Hence,
analysing the stellar velocity field in the framework of a perturbation
model, we can derive both the parameters of the density waves and 
those of the  galactic rotation.

Unlike the above cited papers, let us represent the gravitational potential
$\varphi_G$ of the Galaxy as the sum:
\begin{equation}
\varphi_G = \varphi_0 + \varphi_{S2} + \varphi_{S4}
\end{equation}
where $\varphi_0$ is, as usual, the unperturbed axisymmetric part of the
potential that determines the Galaxy equilibrium as a whole 
($d\varphi_0/dR = \Omega^2R$, $\Omega(R)$ is the angular rotation velocity
of galactic disk, $R$ is the galactocentric distance) and $\varphi_{Sm}$
($m=2$ or $m=4$) are the $m-th$ harmonics of the perturbations due to spiral 
density waves. According to Lin et al (1969):
\begin{equation}
\varphi_{Sm} = {\bf A_m} cos(\chi_m),
\end{equation}
where ${\bf A_m}$ is the amplitude of the $m-th$ harmonic, $\chi_m$ is the
wave phase:
\begin{equation}
\chi _m = m\,\bigl[ \cot ({i_m})\,\ln (R/R_{\odot })-\vartheta \bigr] %
+ \chi _{\odot m} \ ,
\end{equation}
$m$ is the azimuthal wave number, i.e. the number af arms for a given
harmonic, $i_m$ \ is the corresponding pitch angle of the arms, 
$R$, $\vartheta $ are the polar  coordinates with the origin at the Galactic
center, $R_{\odot }$ is the solar galactocentric distance, $\chi _{\odot m}$
is the initial phase or the wave phase at the Sun position. From Eq.(3) it
is seen that this last value fixes the $m-th$ harmonic position relative
 to the Sun.

In accordance with Eq (1) we can write the perturbed stellar velocity for the
radial (directed along galactocentric radius) ${\tilde v}_R$ and the
azimuthal ${\tilde v}_\vartheta$ components as follows:
\begin{equation}
{\tilde v}_R = f_{R2} \cos ({\chi _2}) + f_{R4} \cos ({\chi _4}) \ ,
\end{equation}
\begin{equation}
{\tilde v}_\vartheta = f_{\vartheta 2}\,\sin ({\chi _2}) + f_{\vartheta 4}%
\sin ({\chi _4}) \ ,
\end{equation}
where \ $f_{Rm}$\ and \ $f_{\vartheta m}$\ are the amplitudes of $m-th$
harmonic. These quantities are related to the parameters of the density
waves  by
some formulas (see Lin et al, 1969).

Substituting Eqs (4-5) into Eqs.(1,2) of MZ and using
the statistical method described by  MZ and MZDMR, we derive the parameters 
of the rotation curve ($\Omega_{\odot },A,R_{\odot }^{}\Omega
_{\odot }^{\prime \prime }$\,) and the parameters of the spiral waves
($f_{Rm}, f_{\vartheta m}$, $i_m, \chi _{\odot m}$) over the observed stellar
 velocity field. Then by means of the density wave theory (Lin et al, 1969)
 we compute the difference $\Delta \Omega_m$  between the angular rotation
 velocity of $m-th$ pattern $\Omega _{pm}$ and the rotation velocity of the
 Galaxy at the Sun position ($\Delta %
\Omega _m = \Omega _{pm} - \Omega _{\odot }$). By equating
$\Omega (R_{cm}) = \Omega _{pm}$  we find the displacement $\Delta R_m$
of the Sun relative to the corotation radius $R_{cm}$ ($\Delta R_m = R_{cm}
 - R_{\odot }$, for computational details see  MZ and MZDMR).

Notice here some features of our task. It is widely believed that 
the spiral structure of our Galaxy is generated by a bar in the galactic
center (e.g., Marochnik \& Suchkov, 1984). So, the angular rotation velocity
of the pattern is determined by the bar rotation. Hence, the value
$\Omega _{pm}$ does not depend on $m$, and the corotation radius does
not depend on $m$ either.

Further, the nature of the perturbed gravitational field could be two-fold.
First, in Lin et al.(1969) theory, the spiral waves are self-sustained and
the disturbances of the gravirational field are mainly due to the density
waves which propagate in the galactic disk. The corresponding dispersion
relation imposes a connection between $i_2$ and $i_4$.
Indeed, the solution of the dispersion relation for spiral
waves relative to the radial wave number $k_m$ in the vicinity of corotation
does not depend on $m$ (Shu, 1970, Mark, 1976).
Since $\cot (i_m) = k_{m}R/m$, we have
\begin{equation}
\cot (i_2) = 2\cot (i_4)\,.
\end{equation}
Therefore, the two-armed pattern is tighter
wound (about twice for small pitch angles) than the four-armed one.

The above argument is strictly held in the vicinity of the
corotation circle (according to Cr\'ez\'e \& Mennessier, 1973; MZ and MZDMR;
AL, etc. the Sun is situated just near
the corotation). However, since the radial wave number and consequently the
pitch angle are slowly varying functions of $R$ (Lin et al, 1969) the
relation (6) can be used for a sufficiently wide region around the corotation
radius.

In the second approach the perturbations of the gravitational potential are
mainly due to a  bar in the galactic center. According to the models in
which 2 of the 4-arms components coincide with the 2-arms components
(eg. Englmaier \& Gerhard 1999, Al):
\begin {equation}
\chi_4=2\chi_2; \ \ i_4=i_2.
\end {equation}

In what follows we shall analyze both these approaches. We call them the
self-sustained model (approach 1) and the bar-dominated model (approach 2). 

Another important peculiarity of our task is that some of the wave
parameters: $f_{R2}, f_{\vartheta 2}$, $ f_{R4}, f_{\vartheta 4},
i_2$ ($i_4$ is fixed automatically in the both models), 
$\chi_{\odot 2}$ and $\chi_{\odot 4}$, obey  nonlinear statistics.
In the self-sustained model $\chi_{\odot 2}$ and $\chi_{\odot 4}$ are
considered to be independent quantities; in the bar-dominated  approach, these
values are connected by Eq (7). But if we fix the pitch angles and the
initial wave phases, the task becomes a linear one over other the
parameters. So, the strategy to localize the global minimum for the
residual ($\delta^2$) in the second approach is just the same presented by
MZ and MZDMR. However, in the self-sustained model it slightly
changes. In this case we have to fix 3 independent quantities: $i_2,
\chi _{\odot 2}$ and $\chi _{\odot 4}$, and look for the minimum of
residual over other the parameters by means of the least square method
(we denote this minimum by $\Delta=min_{\,i_2,\chi_{\odot 2},\chi_{\odot 4}}
(\delta^2) $). Then we change values of $i_2,
\chi _{\odot 2}$ and $\chi _{\odot 4}$ and again derive $\Delta$, and so on.
After that, we construct the net $\Delta$ as a function of $i_2,
\chi _{\odot 2}$ and $\chi _{\odot 4}$. Of course, we cannot imagine this
function visually, but we can construct the surface, say,
$\Delta (\chi _{\odot 2}, \chi _{\odot 4})$ for a set of values $i_2$
and check by eye  localization of the global minimum of the residual.
After the minimum is localized, by means of a linearization
procedure (Draper \& Smith, 1981) we define more exactly the searched
 parameters and the covariation matrix of errors (see details in MZ).


\section{Results of Cepheid kinematics analysis}

The above described procedure was applied to a sample of Cepheids which
represent the best observational material for solving the above formulated
problem. For the Cepheids both the line-of-sight velocities (Pont et al,
1994, Gorynya et al, 1996,  Caldwell \& Coulson, 1987) and proper
motions (HIPPARCOS, ESA, 1997) are available. 
Let us discuss the results derived for the above approaches separately.

{\it Self-sustained model.}
For illustration in Fig.1  the
surfaces $\Delta (\chi _{\odot 2}, \chi _{\odot 4})$ are given for
3 values of $i_2$.One can easily see the global minimum in the vicinity
 of $i_2\approx -6^{\circ}$. The final values of the parameters with their
 errors are given in Table~1.

First of all, we notice the significant decrease of the residual
$min (\delta^2)$ in this case in comparison with the one for the single
harmonic $m=2$  or $m=4$ (see Table 1 of MZ, runs no 3 and 8). 
As it was shown in the cited paper, the inclusion
of the spiral perturbation in stellar motion happens to be significant.
However the authors could not make a choice between the two alternatives:
pure 2- or 4-armed patterns.

Now by means of F-test we can show that the previous hypotheses
that the pattern can be represented by only one harmonic ($m=2$ or $m=4$), 
should be rejected in favor of the alternative hypothesis that the pattern
 is well represented by superposition of 2+4-armed pattern. In other words, the
representation of the galactic structure by a superposition of 2+4-harmonics
is clearly preferable to the one that uses a single harmonic.

The quantities $\Delta \Omega_m$ and $\Delta R_m$ can both be calculated
from the parameters obtained for $m=2$ and for $m=4$. It is not hitherto obvious
that for different $m$ the quantities occur to be the same as we have
supposed above. But our calculations lead to very close values for the
corresponding quantities:
$\Delta \Omega_2 = 0.15 \: km\, s^{-1} kpc^{-1}$
and $\Delta \Omega_4=0.18 \: km\, s^{-1} kpc^{-1}$;
accordingly $\Delta R_2$=-0.03$ \,kpc$ and
$\Delta R_4$=-0.04$ \,kpc$,
the standard (i.e. 68\%) confidence intervals being for
$\Delta \Omega_2:\, -0.61$ to $1.02\, km\,s^{-1}kpc^{-1}$;
for $\Delta \Omega_4:\, 0.13$ to $0.24 \, km$ $s^{-1}kpc^{-1}$;
for $\Delta R_2: \, -0.21$ to $0.13 \, kpc$;
for $\Delta R_4: \, -0.05$ to $0.03 \,kpc$
(the confidence intervals were estimated by means of numerical experiments
described by MZ). Hence, in the model under consideration the Sun is
practically situated at the corotation circle, slightly beyond it.

We cannot estimate with any reasonable accuracy the values of the
amplitudes of the spiral gravitational field $\bf A_2$ and $\bf A_4$
(this is expected from linear perturbation theory). However, their
ratio is derived very precisely: ${\bf A_2/{\bf A_4}}=0.79$, 
the standard confidence interval being $0.77$ to $0.80$.

 The locus of minima for $\varphi_{Sm}$ are shown in Fig. 2; they are
 the lines of constant phase $\chi_m$ on the galactic plane corresponding to 
$min \,\varphi_{Sm}$. From this figure the pattern may be thought
to be 6-armed one. However, this is not the case in our model, in which the
potential perturbation is represented by a cosine functions. Indeed, simple
 computation shows that for the above derived parameters the
 sum $\varphi_{S2} + \varphi_{S4}$ has at most 4 minima over $\vartheta$
 for a fixed $R$. It would only be possible to obtain 6 minima (in the frame
of a 2+4 armed model)  if the potential were represented by some function
 presenting  sharp minima, contrary to the cosine function. 
The visible structure derived by means of particle-cloud simulations and
given in Sec. 4 supports this point of view. Of course, the actual
structure of the Galaxy may occur to be more complicated e.g., due to
 higher wave harmonics or to special effects at the corotation (Mark, 1976).
 However, these possibilities are beyond the present investigation.

{\it Bar-dominated model.}
In this case the results are quite different.
The pattern rotation velocities are: for $m=2$ 
$\Omega_{p2} = 35.0 \ km s^{-1} kpc^{-1}$ and for $m=4$
$\Omega_{p4} = 29.2 \ km s^{-1} kpc^{-1} $. 
So that the main requirement of the model
(independence of pattern rotation velocity and the corotation radius from
$m$) is not held. Indeed, for parameters of rotation curve of Tables 1 and 
the above value for $\Omega_{p2}$ the corotation radius
$R_{c2}$ does not exist as a real number at all 
(this is mainly because the second
derivative $R_{\odot}\Omega^{''}_{\odot}$ is negative). 
On the other hand for $m = 4$ $R_{c4} = 5.4 \ kpc$. Further, 
in this case the ratio ${\bf A_2 / \bf A_4} \approx 8.21$ is very large.
So, the visible pattern happens to be 2-armed!

 At last, the residual
$\delta^2 \approx 220$ is significantly greater than in the previous
approach and occurs to be very close to the values for pure $m=2$ and pure
$m=4$ solutions (see Table 1 of MZ 1, runs no 3 and 8). Hence, it is
impossible to make a choice between pure
$m=2$, pure $m=4$ or superposition of 2+4-modes in the bar-dominated approach.

The above statistical analysis of the large-scale stellar kinematics of
Cepheid stars,  leads us to the conclusion that the preferable solution for
the spiral structure of the Galaxy is a superposition of {\it self-sustained} 
2+4-harmonics of density waves, and that the Sun is situated very close to
the corotation circle.

\section{Visible large-scale structure of the Galaxy}

In the previous Section we derived the structural parameters for the
gravitational field of the Galaxy. However, as it was mentioned above, it is
not directly seen. To make visible the above derived structure, 
the evolution of a gas cloud ensemble in the galactic gravitational field
pertubed by spiral arms  will be next  considered. Following Roberts
 \& Hausman (1984), we simulate the
interstellar clouds by ballistic particles moving in a given gravitational
field with a potential $\varphi_G$ defined by Eqs (1-3). Let us briefly
describe the formulation of the task.

At the initial moment of time ($t=0$) the spiral perturbations are 
assumed to be absent ($\varphi_{Sm} = 0$). $N$ particles
($N = 2\cdot 10^4$) are uniformly distributed over a disk within
$R < 13 \, kpc$. Each particle is given the local rotation velocity, disturbed
 by a chaotic velocity with one-dimensional dispersion $8 \ km \ s^{-1}$.
For $t>0$ the spiral perturbation is "switched on". Our task is to compute
the reaction of the system on this perturbation (the $N$-body problem for
particles moving in an external field).

The parameters both for unperturbed and for perturbed potential were taken
from Table 1. Since the derived rotation curve is valid in a restricted
region, for $R > 9.4 \ kpc$ we continue it by a flat part.

The self-sustained galactic waves are well known to exist between the inner
and outer Lindblad resonances. Since we do not take into consideration
a bar in the galactic center, the spiral perturbation was cut off for $R
< 2 \ kpc$. For the spiral gravitational amplitude $\bf A_2$ we assumed the
"standard" value: $2{\bf A_2} cot(i_2)/\Omega{^2}_\odot R{^2}_\odot = 0.05$
(Lin et al, 1969).
In the simulation, collisions between the interstellar clouds occur, the
 collisions being energy-dissipating. For the computation of this process,
 we used the method described in detail by Roberts \& Hausman (1984). For
 the cloud cross-section  we adopted a typical value for the H I clouds.
 Mutual clouds gravitation, or other effects like interaction of the
 clouds with expanding envelopes of supernovae, etc., were not taken into
 account. We also restrict ourselves to
consideration of two-dimensional particle motion in the galactic plane. All
computations are performed in a frame of reference corotating with spiral
arms. 

The result of our  simulation of particle-cloud dynamics in the spiral
 gravitational field  for the superposion of 2+4 self-sustained density
 wave harmonics is shown in Fig. 3. There is
good agreement in major features with Fig. 3 of Efremov (1998). In a
significant part of galactocentric distances the pattern looks like a 4-armed
one. But we do not face with the problem of too short arms, as it would be
in the case of pure $m=4$ harmonic (see AL). Our pattern reflects the
complicated picture often observed in external galaxies, 
e.g. arm bifurcation or their overlaping.

\section{Gap in the galactic gaseous disk as an indicator for the corotation
circle}

One of the most important conclusion of Sec. 3 is that the Sun lies very
close to the corotation radius. In this Section we present a new test which
turns possible to localize directly the position of the corotation
circle in a spiral galaxy.

Many years ago Kerr (1969) payed attension to a ring-like region which is
markedly deficient in neutral hydrogen, with radius  slightly greater
than the solar distance from the galactic center (see also Simonson, 1970).
This result  was later supported in more detail by Burton (1976). He showed
that there is a very clear gap in radial distribution of atomic hydrogen
in our Galaxy at $R \approx 11 \ kpc$, whereas in the old scale, used in
that paper, $R_{\odot} = 10 \ kpc$ (see Fig. 6 of Burton, 1976). 
In general, the gap reminds the Cassini gap in Saturnian ring.

It is natural to connect this gap in the radial ISM distribution with the
process occuring at corotation: the gas is pumped out from the corotation
under the influence of the gravitational field of spiral arms
(see also Suchkov, 1978). The qualitative explanations is as follows. It
is well known, that when the gas flows through the galactic density waves,
a  shock arises in the medium (Roberts, 1969; Roberts \& Hausman, 1984).
Since the galactic disk rotates differentially and for $R < R_c$
 \ $\Omega > \Omega_p$, the gas in this region overtakes the spiral wave,
entering it from the inner side. In the shock the clouds are decelerated,
and fall towards the galactic center. For $R > R_c$ \ $\Omega < \Omega_p$,
and the process is inverse. Here the wave overtakes the gas and pushes it.
So, the clouds pass to an orbit more remote from the galactic center
(see also Goldreich \& Tremaine, 1978 and Gor'kavyi \& Fridman, 1994). 

Our simulation of gas cloud dynamics in spiral gravitational field 
directly demonstrates this phenomenon. We show in Fig.4  the radial gas
distribution ($<n>$ is the particle concentration averaged over a circle)
for $t = 0$ and $t = 3.0$ (the time is given in rotation period at the solar
distance) for the best parameters of Sec 3. The gap in the ISM
distribution at the corotation radius is well seen. Comparison of  Fig.4
with  Fig. 6 of Burton (1976) shows a close similarity between them.

This result enables us to explain another problem as well. It is
well known that the rotation velocity of the  disk presents a sharp minimum
near the solar galactocentric distance; the minimum appears independently
of the tracer being gaseous (eg. Honma \& Kan-ya, 1998) or stellar (Amaral
et al, 1996). This phenomenon, could be thought to be understood in terms of
the velocity perturbation from the galactic density waves, or from the
rising effect of a dark matter component at distances larger than the solar
radius. Amaral et al. exclude these hypotheses, in their discussion of
the nature of this minimum. Now, if there is a ring-like region devoid of gas,
in principle the rotation velocity could not be measured in that region,
using a gas tracer. Similarly, if one selects short-lived stars as tracers,
these stars are not expected to form and to exist inside that region,
because of the gas deficiency and of the very low velocity of the gas with
respect to the spiral pattern, which turns the gas compression (the
star-formation process) inefficient. It is therefore probable that the gas
clouds and the stars that we observe  inside the gap are objects with
non-circular orbits that invade the gap; they are observed close to their
maximum elongation and therefore, present smaller velocity than the
circular one, in the direction of rotation.

So, a ring-like gap in the galactic gaseous disk, and possibly also
a sharp minimum in the rotation curve, may serve as an independent indicator
for localization of the corotation circle in a spiral
galaxy.

\section{New spiral structure of the Galaxy derived from H II data}

The H II regions are the best tracers of the large-scale spiral structure,
since they can be observed at large distances, and unlike H I, they are
sharply concentrated in the arms (e.g. GG). We
performed a new analysis of the sample of H II regions of the Galaxy,
without making use of the results of the theoretical models discussed in
this paper, except for one hypothesis, namely, that the structure can be
represented by a superposition of 2+4-armed patterns. Therefore, the results
of this section constitute an independent test of the previous ones.

The procedure adopted is to trace spiral arms in the galactic plane, and to
transform $X - Y$ positions along the arms into locus of the arms in the $l-v$
diagram, by means of the rotation curve. By varying the parameters of the
arms, we looked for the best fit to the $l-v$ diagram. It is well known that
the arms that are situated inside the solar circle, transform into narrow
loops in the $l-v$ diagram; the extremity of a loop corresponding to a
tangential direction in the galactic plane. 
The observed pattern is represented by the sum of the $m = 2$ system (two
identical long arms with phase difference of $180\arcdeg$) and of the $m = 4$
system (four identical short arms each separated by $90\arcdeg$ in phase)
shifted by some phase angle from the first system. The
adjusted parameters are the pitch angles, the angle for the phase shift, the
inner and the outer radii of each of the two systems.

We used  the catalog of  H II regions of Kuchar \& Clark (1997). The $l-v$
diagram  with the fitted loops, is shown in Fig.5. The rotation curve,
that we adopted for the position-velocity transformation was derived from
the interstellar gas data of Clemens (1985), reinterpreted in terms of
$R_0 = 7.5 \,kpc$. In this transformation we also took into account a
non-circular motion, due to the fact that when they form, H II regions have a
systematic velocity of about $10 \, km \,s^{-1}$ towards the center of the
Galaxy, that keep them in orbits which are close to the spiral arms
(Bash, 1981). Note that the choice of this parameter considerably helps
in obtaining good fits of the observed locus of the inner  HII regions. The
 loops are labeled {\it a} to {\it i} in the $l-v$ diagram; the corresponding
positions of the tangential points are indicated on the pattern 
in the galactic plane, in Fig. 6.

We next discuss some of the features of the proposed structure. In the region 
$l$= 340$\arcdeg$ to 270$\arcdeg$, which is the only region where clear arms
were observed by GG,  the  structure resulting from our study closely
resembles that of these authors, presenting about the same tangential
directions. Remark that the longitudes of tangential directions are not
affected by a change of distance scale (GG used $R_0$ = 10 kpc). Therefore, 
in this range of longitude, it is almost impossible to distinguish between
the 2+ 4 arms model that results from our study, and the empirical model
of GG.This is specially true if we adopt a smooth function to represent
the potential, like we did in previous sections, since close potential
minima tend to merge. A difference between GG and our work is that  
 GG did not indicate the existence of a tangential direction at about
 $l$=338$\arcdeg $(our inner loop $e$), but obviously there are observed
HII regions in that direction, at large negative velocities, that justify
our model. In some other directions, the observations favor our model as well.
Remark for instance that there are concentrations of H II regions near
labels {\it a} and {\it f} in Fig. 5. These are well explained by a spiral
arm that pass very close to the Sun, seen almost at $l \approx +90\arcdeg$
and then at $l \approx -90\arcdeg$, since the velocities are almost zero
 in these directions, for distances that are not too large, according to
 the well known expression $v \propto sin (2l)$.
The longitudes of the tangential directions indicate that it is an arm with
small pitch angle (about $6\arcdeg$). On the contrary, the wide loop label
 {\it b}, can only be well reproduced  with a larger pitch angle
 ($12\arcdeg$).This emphazises the nedd for arms with different pitch angles.
 The best fit of the $l-v$ diagram of the observed HII regions that we
 obtained with a simple 2 + 4 armed model, although not perfect, reproduces
 the main features  of the diagram, and in particular,
the main tangential directions. This was obtained with  pitch angles 
$6.6\arcdeg $ for the 2-armed component and $12\arcdeg$ for the 4-armed
 component.

Let us now discuss the theoretical $l-v$ diagrams, that were computed by means
of our particle simulations in Sec. 4 and are shown in Figs. 7 to 9. Figs.7
and 8 are very similar to the observed diagram for H II regions, in many
aspects. In particular, the observed loops like {\it {b, c, d, e}} which are
interpreted in Fig. 5 as empirical locii of spiral arms, with  corresponding
tangential directions to the arms, can be clearly seen in Fig. 7 and 8 as
well. A difference that appears between the theoretical $l-v$ diagrams and
the observed ones for H II regions is a strong concentration of points along
the line from $l \approx - 60^{\circ}, \,v \approx + 100 \,km \, s^{-1}$ to 
$l \approx + 60^{\circ}, v \approx -100 \, km \,s^{-1}$. This is an expected
result since the points corresponding to H I (recall that the particles
simulate the H I clouds, see Sec 4) are situated at large galactic radii, 
where H I is known to exist, whereas there is lack of H II regions at large
 radii.

It is not surprising that the $l-v$ diagrams in Figs. 7 and 8 are so similar,
since both models contain similar 2-arms component, and this component
 is prominent over a wider galactocentric region than the 4-arms component
 of the 2+4 arms model. However, we can point out a number of differences.
 For instance, the observed loop $b$, which is due to a 4-arms component,
is correctly reproduced by the particles of the self-sustained model 
(Fig. 7) and does not appear in Fig.8. On the other hand, the arm that passes 
very close to the Sun, as discussed above (loops $a$ and $f$) appear more
clearly in the 2-arms model (Fig.8). We emphazise that the loops presented
in Figs. 7-9 are eye-guides for comparison of theoretical diagrams with
observed HII regions, but they are not perfect fits to the HII regions.
In the HII regions diagram (Fig. 5) we can see many objects between loops
$c$ and $d$, that are not well fitted by the loops. We can see many objects
as well in this region in Figure 8, the theoretical self-sustained model.
In other words, the theoretical model is closer to reality than the
``empirical'' fit represented by the lines. Around longitude 240$\arcdeg$
(or -120$\arcdeg$), the theoretical $l-v$ diagram of the self-sustained
shows many objects with velocities of the order of 30 kms$^{-1}$. These
objects seem to delineate a spiral arm that is not seen in the HII regions
diagram. However, a arm indeed exists at this position, as can be seen
in the longitude-velocity diagram of IRAS sources given by Wouterlout et al.
(1990). This arm is deficient in HII regions, probably due to the proximity
of corotation.

On the contrary, the 4 arms model (Fig. 9) show stronger differences with
observations. The particles do not show loops $a$ and $b$, but show a clear
loop between loops $c$ and $d$, that is not present in the HII regions diagram.

In summary, the pure 2-arms model and the self-sustained 2+4 arms model
produce theoretical $l-v$ diagrams that are  similar between them and
similar  to that of observed HII regions. The only striking difference
between the theoretical and observed diagrams is an expected one, due to
 the fact that we are comparing objects that behave differently (HI gas
and HII regions). If we look into the details,  the 2+ 4 arms self-sustained
model is favored. The choice in favor of the 2+4 arms model, compared to
the pure 2-arms, is more strongly dictated by the study of Cepheid
kinematics.  


\section{Conclusion}

In the present research a new approach to the problem of the galacic spiral
structure was proposed in order to construct a more realistic picture like
often seen in external galaxies: co-existing of different spiral systems in a
galaxy, arm bifurcation and their overlaping. Our theoretical considerations
shows that superposition of self-sustained spiral wave harmonics could
explain some of the above features, since different azimuthal wave harmonics
have different pitch angles.

In the framework of the simplest model of superposition of 2+4-armed spiral
wave harmonics, we analysed the best up-to-date data on stellar kinematics,
which is the sample of Cepheid stars, with proper motions and parallaxes
determined by HIPPARCOS. We examined two models, the self-sustained and
the bar-dominated waves. 
This study complements the previous studies of pure 2-arms and pure 4-arms
presented by MZ ad MZDMR. Of the four models, clearly the one that gives
the best fit to the Cepheid kinematics is the self-sustained model, which
is a superposition two arms with pitch angle about 6$\arcdeg$ and four arms
 with pitch angle about 12$\arcdeg$. We  performed  N-particle simulations to
make visible the structure of the  potential derived from the Cepheid
kinetics for the four models, and to construct the corresponding $l-v$
diagrams.

As an independent test of a 2+4 arms model,we performed a new analysis
of the $l-v$ diagram of the galactic HII regions, which
are the best tracers of the large-scale spiral structure. We fitted the
 observed $l-v$ diagram empirically with the locii of spiral arms, using
 a 2+4 arms model.  Coincidentally, the best empirical fit was again found
 using pitch angles about $6\arcdeg$ and $12\arcdeg$. We also compared
the theoretical $l-v$ diagrams from the particle simulations with the locii
 of arms derived from  HII regions.Although the differences between $l-v$
diagrams of the  theoretical models (pure 2-arms, pure 4-arms, and 2
different models of 2+4 arms) are not striking, the 2-arms model and the
 self-sustained 2+4 arms models are the ones that produce 
theoretical $l-v$ diagrams most similar to that of the HII regions.

Of all the arguments that we examined, the significantly better fit of the
kinematics (the $\delta^2$ analysis) of Cepheids with the 2+4 armed,
self-sustained model, is the most convincing one, but clearly the analysis
of the HII regions sample gives support to our interpretation. Although
the Galaxy probably shows some deviations from any simplified model, the
2+4 armed model with different pitch angles is the one that constitutes
the best approach, being consistent with observations and with spiral waves
theory. This is the simplest model that can be proposed, apart from pure
harmonic modes. Our model does not exclude the existence of higher modes,
that are allowed to exist near corotation, but these modes are probably 
 less significant than the first harmonics. It is
interesting to remark that our model is able to reconcile  the first model
of Lin \& Shu (1964), which has 2 arms with pitch angle 6$\arcdeg$ similar
to our 2-arms harmonic, with the need to satisfy Kennicut's (1982) correlation
between pitch angle and maximum rotation velocity, which predicts a pitch
angle of about 14$\arcdeg$ for our Galaxy, not very different from that
of our 4-arms harmonic.

As a by-product of the study, our particle simulation shows that a deficiency
of intertellar gas must occur near corotation. This explains the gap in the HI
ditribution observed by Kerr(1969) and by Burton (1976). This effect is also
possibly related to the sharp minimum in the rotation curve near the solar
radius discussed by Amaral et al. (1996), also seen in the curve by Honma
 \& Kan-ya (1998). In external galaxies the ring-like gap in H I distribution
 may directly show the localization of the corotation circle.

Our investigation of Cepheid kinematics points out that the Sun lies
very close to the corotation circle. This conclusion is  supported by
the solar closeness to the ring-like region of H I deficiency in the Galaxy
derived by Kerr (1969) and by Burton (1976), as well as by the position
of Lindblad resonances (as discussed in Section 1) and by direct measurement
of the pattern speed using open clusters (AL).

\vspace{1.0cm}${\it Acknowledgements.}$\
Authors are grateful to the Hipparcos
astronomical team, who presented us the Hipparcos catalogue.


\newpage
{\bf Figure captions}
\smallskip

\noindent Fig.1.- Surfaces of the $\Delta$ as a function of $\chi_{\odot2}$
and $\chi_{\odot4}$ for 3 values of pitch angle $i_2$.

\noindent Fig.2.- The locus of  $min \,\varphi_{S2}$ and $min \,\varphi_{S4}$
on the galactic plane. The scale is indicated in kpc. As discussed in the text,
this does not correspond necessarily to the visible structure.

\noindent Fig.3.- The visible structure of the Galaxy derived for the best
model (superposition of 2+4 self-sustained wave harmonics) by means of
particle-cloud simulation.

\noindent Fig.4.- The radial distribution of cloud concentration $<n>$
(averaged over a circle) for $t=0$ (dashed line) and $t=3.0$ (solid line).
The gap in the gas distribution is clearly seen near the corotation radius
$R_c$. The Sun is situated at $R_\odot = 7.5 \, kpc$.

\noindent Fig.5.- The observed $l-v$ diagram for H II regions (from Kuchar \&
Clark, 1997) and the loops that we fitted empirically with a 2+4 arms
structure.

\noindent Fig.6.- The spiral structure of the Galaxy derived from H II data.

\noindent Fig.7.- The theoretical $l-v$ diagram computed by means of
particle-cloud simulation for the best model: superposition of 2+4
self-sustained wave harmonics..The lines represent the fit to observed
HII regions, from Fig. 5, for camparison.

\noindent Fig.8.- Same as in  Fig. 7, but for the model of pure
$m=2$ wave harmonic (the parameters were taken from MZDMR).The lines
represent the fit to observed HII regions, from Fig. 5, for camparison,
 and not a fit to the $m=2$ model.

\noindent Fig.9.- Same as in  Fig. 8, but for the model of pure
$m=4$ wave harmonic (the parameters were taken from MZ)..The lines
represent the fit to observed HII regions, from Fig. 5, for camparison,
and not a fit to the $m=4 model$.

\newpage

\tabcolsep=5pt    
\begin{table*}
\caption{Model parameters and their errors
derived by means of statistical analysis.}

\footnotesize
\begin{tabular}{@{}crrcccccccccccc@{}}
 \noalign{\smallskip}
 \hline
 \noalign{\smallskip}

  $appro-$ &
  $ {\Omega_{\odot}}$
& $ {A}\ \ $
& $ {R_{\odot}\Omega''_{\odot}}$
& $ {u_{\odot}}$
& $ {v_{\odot}} $
& {\normalsize $\left| i_2\right| $}
& $ {\chi _{\odot 2}}\ $
& $ {f_{R2}} $
& $ {f_{\vartheta 2}}\ $
& {\normalsize $\left| i_4\right| $}
& $ {\chi _{\odot 4}}\ $
& $ {f_{R4}} $
& $ {f_{\vartheta 4}}\ $
& min ${\delta^{2}} $ \\

$ach$ &
  $ (\frac {km}{s\,kpc}) $
& $ (\frac {km}{s\,kpc}) $
& $ (\frac {km}{s\,kpc^{2}}) $
& $ (\frac {km}{s}) $
& $ (\frac {km}{s}) $
& $ ({}\arcdeg) $
& $ ({}\arcdeg) $
& $ (\frac {km}{s}) $
& $ (\frac {km}{s}) $
& $ ({}\arcdeg) $
& $ ({}\arcdeg) $
& $ (\frac {km}{s}) $
& $ (\frac {km}{s}) $
& {$ $} \\
\noalign{\smallskip}
\hline
\noalign{\smallskip}

$self-$&$ 26.3  $ & $   17.5  $ & $    9.9  $ & $   -8.8  $ & $   11.9  $ & $   6.1   $ & $   311.  $ & $   0.4   $ & $  -14.0  $ & $  12.0   $ & $   122.  $ & $   0.8   $ & $  -10.9  $ & $187.$ \\
$sust.$&${\pm 1.3}$ & ${\pm 0.8}$ & ${\pm1.9}$ & ${\pm 1.0}$ & ${\pm 1.1}$ & ${\pm 0.4}$ & ${\pm 11.}$ & ${\pm 3.0}$ & ${\pm 3.0}$ & ${\pm 0.8}$ & ${\pm 15.}$ & ${\pm 3.3}$ & ${\pm 2.9}$ &  \\
\noalign{\smallskip}
\hline
\noalign{\smallskip}

$bar-$ & $ 25.7  $ & $    9.8  $ & $   -6.8  $ & $  -13.6  $ & $    9.3  $ & $  12.6   $
& $  184.$ & $ 12.5 $ & $ -19.4 $ & $ 12.6 $ & $ 8. $ & $ 6.6 $ & $ -10.1 $
& $ 220. $ \\
$domin.$ & $\pm 1.2$ & $\pm 2.3$ & $ \pm5.0$ & $\pm 2.4$ & $ \pm 1.2 $ & $\pm 0.5$ &
$\pm 4.$ & $\pm 3.8 $ & $\pm 4.4. $ & $\pm 1.0$ & $\pm 7.$ & $\pm 2.4$ &
$\pm 2.1$    \\ 
\noalign{\smallskip}
\hline

\end{tabular}
\end{table*}

\end{document}